\documentclass[final,5p,times,twocolumn]{elsarticle}

\usepackage{amssymb,url}

\journal{NeuroImage}

\begin{document}

\begin{frontmatter}



\title{Maximally reliable spatial filtering of steady state visual evoked potentials \tnoteref{t1} }
\tnotetext[t1]{This work was supported by NIH Grants RO1-EY018875 and RO1-EY015790.} 

\author{Jacek P. Dmochowski \fnref{label2}}
\author{Alex S. Greaves}
\author{Anthony M. Norcia}
\fntext[label2]{Author to whom correspondence should be addressed: (t) +1 (650) 736-2793, (e) \url{jdmochow@stanford.edu}}

\address{Department of Psychology, Stanford University, 450 Serra Mall, Stanford, CA 94305}

\begin{abstract}
\noindent Due to their high signal-to-noise ratio (SNR) and robustness to artifacts, steady state visual evoked potentials (SSVEPs) are a popular technique for studying neural processing in the human visual system.  SSVEPs are conventionally analyzed at individual electrodes or linear combinations of electrodes which maximize some variant of the SNR.   Here we exploit the fundamental assumption of evoked responses -- reproducibility across trials -- to develop a technique that extracts a small number of high SNR, maximally reliable SSVEP components.  This novel spatial filtering method operates on an array of Fourier coefficients and projects the data into a low-dimensional space in which the trial-to-trial spectral covariance is maximized.  When applied to two sample data sets, the resulting technique recovers physiologically plausible components (i.e., the recovered topographies match the lead fields of the underlying sources) while drastically reducing the dimensionality of the data (i.e., more than $90\%$ of the trial-to-trial reliability is captured in the first four components).  Moreover, the proposed technique achieves a higher SNR than that of the single-best electrode or the Principal Components.  We provide a freely-available MATLAB implementation of the proposed technique, herein termed ``Reliable Components Analysis''. 

\end{abstract}


\end{frontmatter}


\section*{Introduction}

\noindent When presented with a temporally periodic stimulus, the visual system responds with a periodic response at the stimulus frequency (and its harmonics).   The resulting steady steady visual evoked potential (SSVEP) \cite{kamp1960cortical,regan1966some,tweel1965human,tweel1969signal}  is readily measurable via electroencephalography (EEG) and has been used extensively to probe the spatiotemporal dimensions of visual sensory processing in the human brain (see \cite{regan1989human,vialatte2010steady}).   In addition to their employment in cognitive \cite{andersen2012bottom} and developmental/clinical neuroscience \cite{almoqbel2012technique}, SSVEPs have also been commonly applied to the development of brain-computer-interfaces (BCIs \cite{middendorf2000brain},\cite{ming1999eeg},\cite{zhu2010survey}).     

The volume conduction inherent to EEG spatially smoothes the electric currents generated by cortical sources, thus lowering the spatial resolution of the resulting scalp measurements.  Viewed in another manner, however, the spatial diversity brought about by volume conduction means that the underlying neural signal may be picked up across multiple locations, each with generally different noise statistics.  Consequently, modern-day SSVEP paradigms employ multichannel recording arrays and afford the experimenter with high-dimensional data sets spanning the electrode montage.  The conventional procedure is to \emph{a priori} select one or a few target electrodes and then to analyze the evoked data in the space of the chosen subset.  In contrast, spatial filtering approaches \cite{garcia2011optimal} exploit the spatial redundancy inherent to EEG and form linear combinations of the data, yielding signal ``components''.   A variety of approaches to computing the spatial filter weights have been proposed: maximizing statistical independence \cite{wang2006practical}, maximizing the variance explained \cite{pouryazdian2009detection}, minimizing the noise power \cite{friman2007multiple}, and maximizing the signal-to-noise ratio (SNR) \cite{blankertz2008optimizing}, \cite{friman2007multiple}.  The spatial filtering approach to SSVEPs yields a parsimonious, low-rank representation of the experimental data with the SNR of the components generally exhibiting an increase over that of individual electrodes.  Moreover, the topography of weights comprising the linear combination can potentially inform one of the (at least approximate) location of the underlying neuronal generators.   

However, this latter potential has not been fully realized by existing spatial filtering approaches \cite{garcia2011optimal}.  The predominant criterion being optimized in current spatial filtering paradigms is the SNR, which increases exponentially with decreasing noise power.  The SNR maximizing approaches thus often operate by steering the array orthogonal to the noise subspace, without controlling for the ensuing signal distortion.   Consequently, the topographies of the resulting components do not always bear resemblance to the scalp projections of actual cortical sources, and are thus difficult to interpret (see Figure 8 in \cite{blankertz2008optimizing}, for example). 

The primary application of spatial filtering SSVEP techniques has been the BCI \cite{bin2009online,garcia2011optimal,lin2006frequency,nam2013common,pan2011enhancing,wang2008brain,zhang2013l1}, where SNR optimization, rather than faithful signal representation, is the primary goal.  By contrast, cognitive and neurobiological imaging research commonly employs SSVEPs to elucidate neural information processing, typically examining the SSVEP at a single electrode or by its (raw unfiltered) topography across the electrode array  (see, for example, \cite{herrmann2001human,kemp2002steady,morgan1996selective,muller2006feature,silberstein1995steadyb,srinivasan2006meg}).

Recently, a novel spatial filtering technique which maximizes the inter-subject correlations among a set of continuous EEG records has been proposed \cite{dmochowski2012correlated}.  This method projects the data of multiple subjects onto a common space such that the resulting projections capture the neural responses common to all viewers.   Here, we adopt a similar approach in the SSVEP context by focusing on the across-\emph{trial} correlations.    The technique exploits the fundamental assumption of evoked responses -- reproducibility across trials -- to identify spatial components of the SSVEP which exhibit maximal trial-to-trial covariance.  In other words, we project the data into a space in which the reliability of the SSVEP Fourier coefficients is maximal.  
The proposed technique operates on single-trial SSVEP spectra and explicitly represents the trial dimension.  This is in contrast to existing component analysis techniques which ``stack'' or concatenate the trial dimension in order to achieve a two-dimensional data matrix (space-by-time or space-by-frequency) from which covariance matrices are typically formed (notable exceptions include \cite{zhang2011multiway,zhang2013l1} which employ a tensor formulation of the data in conjunction with multiway CCA to perform trial selection).   Note that such stacking throws away the structure of evoked response data.  Here, we instead use the third (trial) dimension to focus the spatial filters onto features which are reliably evoked in each trial.  

We apply the technique to one simulated and one real SSVEP data set, and from each extract components that exhibit behavior consistent with physiology (for example, the method recovers dipolar topographies which contralateralize with the stimulated visual hemifield).   Moreover, the SNR of the captured components is significantly higher than that of the ``best'' (i.e., highest SNR) individual electrode.   We contrast the method to both Principal Components Analysis (PCA) and the Common Spatial Patterns (CSP) technique \cite{blankertz2008optimizing} which explicitly optimizes the SNR, and find that the proposed technique yields favorable tradeoffs between plausibility of components and achieved SNR.  Additionally, the proposed method provides a drastic dimensionality reduction as the number of components required to capture the bulk of the trial-to-trial reliability is shown to be more than an order of magnitude lower than the number of acquired channels.  In summary, the proposed technique yields a compact representation of SSVEP data sets with high-SNR, physiologically plausible components by optimizing the characteristic feature of evoked responses -- reliability across trials.  A MATLAB toolbox which contains source code to implement the technique, herein referred to as ``Reliable Components Analysis'' (RCA) is available at \url{github.com/dmochow/rca}.  

\section*{Methods}
\subsection*{Reliable components analysis}
The following details the extension of the method of \cite{dmochowski2012correlated} to the SSVEP context; namely, we propose a component analysis technique which explicitly maximizes the trial-to-trial spectral covariance of the SSVEP.   The approach is inspired by canonical correlation analysis \cite{hotelling1936relations} and its generalizations to multiple subjects \cite{kettenring1971canonical}, differing in that it uses the same projection for all
data sets.   It is conceptually similar to the ``common canonical variates'' method \cite{neuenschwander1995common}, which is based on a maximum likelihood formulation, as opposed to the generalized eigenvalue problem developed in \cite{dmochowski2012correlated} and herein.

Consider an experimental paradigm in which a stimulus is presented  $N$ times, such that  we have a set of $N$ data matrices $\{  {\bf{X}}_1, \ldots , {\bf{X}}_N \}$ where ${\bf{X}}_n$ represents the neural response during trial $n$.  Specifically, the (mean-centered) rows of ${\bf{X}}_n$ denote channels, with the columns carrying real and imaginary Fourier coefficients across the frequency range of interest (i.e., a three-response-frequency paradigm will have 6 columns in ${\bf{X}}_n$).   

In the following,  let ${\mathcal{P}}_i=  \{ (p_{i}, q_{i}) \} = \{ (1,2), (1,3) , \ldots , (N-1,N) \} $ denote the set of all $P=N \times (N-1) /2 $ unique trial pairs.  We then form the following trial-aggregated data matrices:
\begin{eqnarray}
\label{eqn:agg}
\bar{{\bf{X}}}_1 &=&  \left[ \begin{array}{cccc} {\bf{X}}_{p_1}  & {\bf{X}}_{p_2} & \ldots & {\bf{X}}_{p_P}  \end{array} \right] \nonumber \\
\bar{{\bf{X}}}_2 &=& \left[ \begin{array}{cccc} {\bf{X}}_{q_1}  & {\bf{X}}_{q_2} & \ldots & {\bf{X}}_{q_P}  \end{array} \right].
\end{eqnarray}
We apply a linear spatial filter to the aggregated spectral data 
\begin{eqnarray}
\label{eqn:ywx}
\bar{{\bf{y}}}_1&=&  \bar{{\bf{X}}}_1^T {\bf{w}},  \\ \nonumber
\bar{{\bf{y}}}_2&=&  \bar{{\bf{X}}}_2^T  {\bf{w}},.
\end{eqnarray}
where $^T$ denotes matrix transposition.  The correlation coefficient between the resulting spatially filtered data records is given by:    
\begin{eqnarray}
\label{eqn:rho}
\bar{\rho} = \frac{  \bar{{\bf{y}}}_{1}^T  \bar{{\bf{y}}}_{2} }{  (\bar{{\bf{y}}}_{1}^T  \bar{{\bf{y}}}_{1} )^{1/2}  ( \bar{{\bf{y}}}_{2}^T  \bar{{\bf{y}}}_{2} )^{1/2}   }.
\end{eqnarray}
Substituting (\ref{eqn:ywx}) into (\ref{eqn:rho}) yields: 
\begin{eqnarray}
\label{eqn:rho2}
\bar{\rho}  = \frac{ {\bf{w}}^T {\bf{R}}_{12} {\bf{w}}  }{ (  {\bf{w}}^T {\bf{R}}_{11} {\bf{w}}   )^{1/2} (  {\bf{w}}^T {\bf{R}}_{22} {\bf{w}}   )^{1/2} } , 
\end{eqnarray}
where 
\begin{eqnarray}
\label{eqn:cov}
{\bf{R}}_{11} = \frac{1}{2 F P}\sum_{i=1}^{2FP} {\bf{X}}_{p_{i}}  {\bf{X}}_{p_{i}}^T   \nonumber \\   
{\bf{R}}_{22} = \frac{1}{2 F P }\sum_{i=1}^{2FP} {\bf{X}}_{q_{i}}  {\bf{X}}_{q_{i}}^T    \nonumber  \\ 
{\bf{R}}_{12} = \frac{1}{2 F P}\sum_{i=1}^{2FP} {\bf{X}}_{p_{i}}  {\bf{X}}_{q_{i}}^T,   
\end{eqnarray}
where $F$ is the number of analyzed frequencies,  ${\bf{R}}_{11}$ and ${\bf{R}}_{22}$ denote within-trial spatial covariance matrices, and ${\bf{R}}_{12}$ is the \emph{across-trial} spatial covariance matrix which captures trial-to-trial reliability.  Note from (\ref{eqn:rho2}) that $\bar{\rho}$ is the ratio of across- to within-trial covariance.  We seek to find the spatial filter $\bf{w}$ which maximizes this ratio:
\begin{eqnarray}
\label{eqn:rhomax}
\arg \max_{\bf{w}} \bar{\rho}.
\end{eqnarray}
It is shown in \cite{dmochowski2012correlated} that assuming ${\bf{w}}^T {\bf{R}}_{11} {\bf{w}}  = {\bf{w}}^T {\bf{R}}_{22} {\bf{w}}  $, the solution to (\ref{eqn:rhomax}) is a generalized eigenvalue problem:
\begin{eqnarray}
\label{eqn:geig}
\lambda ( {\bf{R}}_{11} + {\bf{R}}_{22} ) {\bf{w}} = {\bf{R}}_{12}  {\bf{w}},
\end{eqnarray}
where $\lambda$ is the generalized eigenvalue corresponding to the maximal trial-aggregated correlation coefficient (i.e., the optimal value of $\bar{\rho}$) achieved by projecting the data onto the spatial filter ${\bf{w}}$.   There are multiple such solutions, ranked in decreasing order of trial-to-trial reliability: $\lambda_1>\lambda_2>\ldots>\lambda_D$, where $D= \mathrm{min} \left[ \mathrm{rank} \left( {\bf{R}}_{12} \right) , \mathrm{rank} \left( {\bf{R}}_{11}+{\bf{R}}_{22} \right) \right]$.  The associated generalized eigenvectors, ${\bf{w}}_1,  {\bf{w}}_2, \ldots, {\bf{w}}_D$ are \emph{not} generally orthogonal.  This is in contrast to PCA which yields spatially orthogonal filter weights.  However, the component spectra recovered by the various $\bf{w}$'s are mutually uncorrelated \cite{golub2012matrix}.    

It is also worthwhile to point out that the assumption ${\bf{w}}^T {\bf{R}}_{11} {\bf{w}}  = {\bf{w}}^T {\bf{R}}_{22} {\bf{w}} $  does not limit generality, as one can simply define ${\mathcal{P}_i}'= \{ (p_i,q_i) , (q_i,p_i) \} = \{ (1,2), \ldots, (N-1,N), (N,N-1), \ldots, (2,1) \} $ and then substitute ${\mathcal{P}_i}'$ in (\ref{eqn:agg}) to ensure that ${\bf{R}}_{11} = {\bf{R}}_{22} $;  this was performed throughout our analyses.  Moreover, when computing the generalized eigenvalues of (\ref{eqn:geig}), we regularize the within-trial pooled covariance by keeping only the first $K$ dimensions, where $K$ corresponds to the ``knee'' of the eigenvalue spectrum, in the spectral representation of ${\bf{R}}_{11} + {\bf{R}}_{22}$.   For the data sets considered here, $K \approx 10$.   Finally, it will be subsequently shown that the bulk of the across-trial reliability is captured in the first $C$ dimensions, where $C \ll D$.  This fact is responsible for the dimensionality reduction of RCA, which we quantify in the forthcoming results by defining the following measure:
\begin{eqnarray}
\label{eqn:propRelExpl}
\eta (C) = \frac{\sum_{i=1}^C \lambda_i }{\sum_{i=1}^D \lambda_i} ,
\end{eqnarray}
where $\eta (C)$ is termed the \emph{proportion of reliability explained} by the first $C$ RCs.

\begin{figure}[h!]
\centering
\includegraphics[width=95mm]{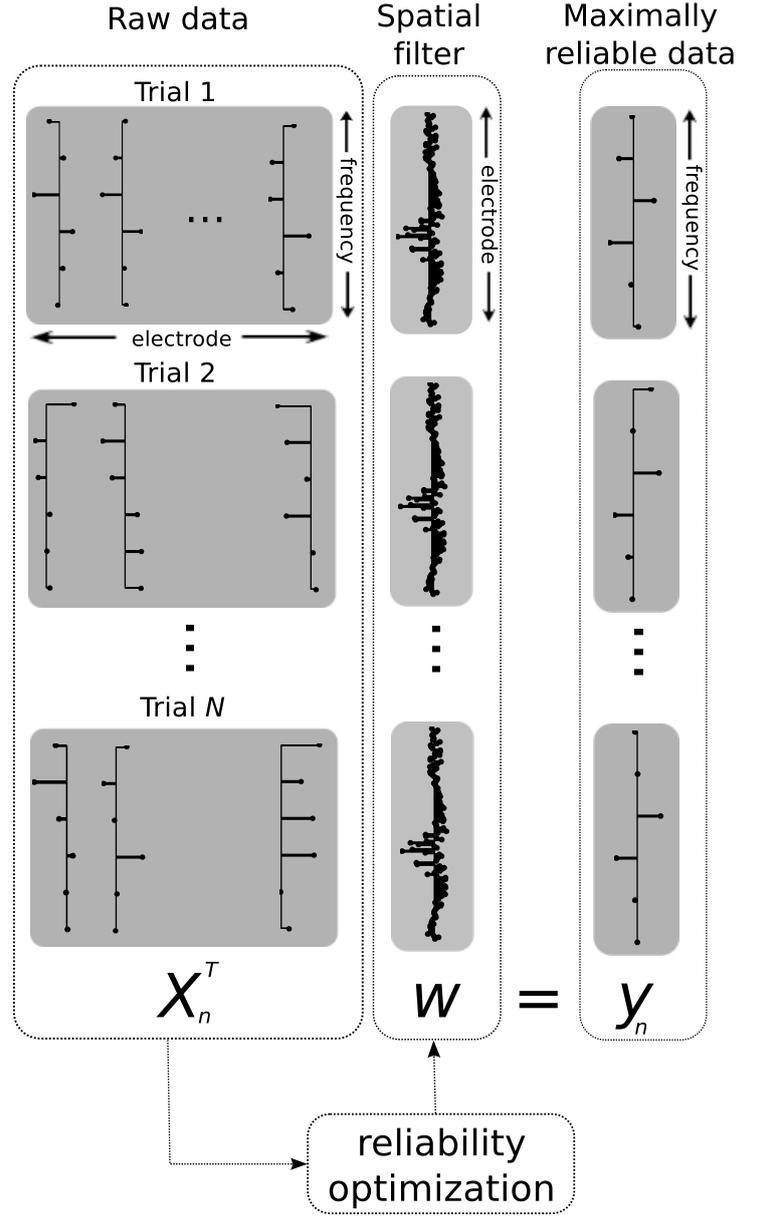}
\caption{\label{fig:process} A diagrammatic view of reliable components analysis (RCA).  A frequency-by-electrode data matrix ${\bf{X}}_n$ which captures the SSVEPs across the array is constructed for every trial $n$.  An optimally tuned spatial filter ${\bf{w}}$ then projects all $N$ such data matrices onto a common space in which the trial-to-trial covariance of the resulting spectra ${\bf{y}}_n$ is maximized.  In our implementation, the raw data matrices ${\bf{X}}_n$ contain the real and imaginary parts of Fourier coefficients at frequencies corresponding to the stimulus frequency and its harmonics.  The projected spectra consist of linear combinations of the spectra at individual electrodes, with the weights of the linear combination chosen to maximize trial-to-trial reliability.   }
\end{figure}

\subsubsection*{Comparison techniques.}
\noindent Throughout the results, we compare the behavior of the proposed technique (shown diagrammatically in Figure \ref{fig:process}) with two of the more commonly employed component analysis techniques: CSP \cite{blankertz2008optimizing} and PCA.   To allow for a fair comparison among these three techniques, care was taken to ensure that all three methods were driven by the same spatial covariances.  To that end,  CSP seeks to project the sensor data onto a space in which the difference between two conditions is maximized  \cite{blankertz2008optimizing}.  In the SSVEP context, these two conditions are simply ``stimulation-on'' and ``stimulation-off''.  As a result, CSP effectively maximizes the following SNR criterion \cite{garcia2011optimal}:
\begin{eqnarray}
\label{eqn:csp}
\max_{{\bf{w}}} \frac{ {\bf{w}}^T  {\bf{R}}_{x} {\bf{w}}  }{ {\bf{w}}^T  \left( {\bf{R}}_{x}+{\bf{R}}_{n} \right)  {\bf{w}}},
\end{eqnarray}  
where ${\bf{R}}_{x} $ is the spatial covariance matrix of the observed data during visual stimulation and ${\bf{R}}_{n} $ is the noise-only spatial covariance (i.e., no stimulation).  For the simulated data set (details forthcoming), we estimated this noise covariance by assuming that a noise-only period was available (i.e., we formed the covariance by simply omitting the propagation of the desired signal to the array).  For the real data set, we estimated the noise covariance matrix from the temporal frequency bands directly adjacent (one below and one above) the signal frequencies considered (i.e., the even harmonics of the stimulation frequency).  Meanwhile, for both simulated and real data sets, ${\bf{R}}_{x} = {\bf{R}}_{11} = {\bf{R}}_{22}$; that is, the within-trial spatial covariance used in the RCA optimization is precisely the observed spatial covariance in CSP.  

By contrast, PCA identifies linear combinations of electrodes which maximize the proportion of variance explained in the observed data:
\begin{eqnarray}
\label{eqn:pca}
\max_{{\bf{w}}} \frac{ {\bf{w}}^T  {\bf{R}}_{x} {\bf{w}}  }{ {\bf{w}}^T {\bf{w}}},
\end{eqnarray}   
with the within trial spatial covariance $ {\bf{R}}_{11} = {\bf{R}}_{22}$ taking the place of ${\bf{R}}_{x}$ in the implementation.  Importantly, the same regularization value $K$ (described above) was used for all three methods.  

\subsubsection*{Component scalp projections}
\noindent When comparing component topographies, we contrast not the spatial filter weights yielded by the appropriate optimization problem, but rather the resulting scalp projection of the activity recovered by that spatial filter.   This inverse topography is generally more informative than the weights in that it encompasses both the filter weights as well as the data that is being multiplied by them \cite{haufe2014on}.    Specifically, let us construct a weight matrix $\bf{W}$ whose columns represent the spatial filter weight vectors $\bf{w}$ yielded by a component analysis technique.  The projections of the resulting components onto the scalp data are given by \cite{haufe2014on,parra2005recipes}:
\begin{eqnarray}
{\bf{A}}={ {\bf{R}}_x {\bf{W}} ( {\bf{W}}^T {\bf{R}}_x {\bf{W}}  )^{-1} },
\end{eqnarray}
where ${\bf{R}}_x= {\bf{R}}_{11} = {\bf{R}}_{22}$ is the (within-trial) spatial covariance matrix of the observed data.  The columns of $\bf{A}$ represent the pattern of electric potentials that would be observed on the scalp if only the source signal recovered by ${\bf{w}}$ was active, and  inform us of the approximate location of the underlying neuronal sources (up to the inherent limits imposed by volume conduction in EEG).

\subsection*{Synthetic data set}
\noindent Prior to delving into real data, we evaluated the proposed technique in a simulated environment.  Simulations possess several desirable properties stemming from the fact that one has access to ground-truth signals, which is particularly beneficial for the computation of the achieved SNR.   A simulation also allows one to easily sweep through parameter spaces; here, we perform a Monte Carlo simulation which evaluates the effect of the number of trials on the recovered components. 500 draws were simulated for each of the following number of trials per draw: $\{10, 20, 30, 50, 100\}$.  Our aim was to assess the behavior of the proposed and conventional component analysis techniques as a function of the amount of available data from which to learn the required spatial filters.    

With the availability of detailed, anatomically-accurate head models, the volume conduction aspect of EEG can be readily modeled in a simulation \cite{hallez2007review}.  To that end, we employed a three-layer boundary element model (BEM) model which was accompanied by labeled cortical surface mesh regions-of-interest (ROIs, see \cite{ales2010v1} for details).  We chose one of these cortical surface ROIs as the ``reliable'' source which models the evoked signal, and another ROI as the variable source which models spontaneous activity not reliably evoked by the paradigm.   We subsequently contrasted the topographies of the recovered components with the lead fields of these simulated sources, thus shedding light on the ability of the various techniques to recover the signal generators.  In other words, we probed whether the techniques recover the underlying sources, and if so, under what conditions.

Specifically, the BEM head model consisted of $20484$ cortical surface mesh nodes and $128$ electrodes placed on the scalp according to a subset of the 10/5 system \cite{oostenveld2001five}.   As mentioned above, two sources were modeled:   the reliable source was designated as the set of all nodes adjacent to the calcarine sulcus (``peri-calcarine'') and consisted of $232$ nodes.   The SSVEP generated by this source had unit amplitude and a fixed phase angle at every trial.  Meanwhile, the ``variable'' source was located in the lateral orbital frontal gyrus and spanned $531$ mesh nodes.  The phase angle of the SSVEP generated by this source was randomly drawn from a uniform distribution over $(0,2\pi)$, with its amplitude drawn from a zero-mean normal distribution with unit variance. Both reliable and variable sources consisted of two frequencies ($F=2$, modeling two even harmonics, for example).  Additionally, additive white Gaussian noise was added at each electrode, with the variance of the noise matched to the average signal variance across the $128$ electrode array.   The resulting SNR (with the ``noise'' encompassing both the variable source and the additive noise) had a median (across trials) value of $-22$ dB, which is in the estimated range for the real EEG signal \cite{goldenholz2009mapping}.  

\subsection*{Real data set}
\noindent SSVEPs were collected from $22$ subjects  (gender-balanced, mean age 20 years) with normal or corrected-to-normal visual acuity. Informed consent was obtained prior to study initiation under a protocol that was approved by the Institutional Review Board of Stanford University.  Visual stimuli were presented using in-house software on a contrast linearized CRT monitor with a resolution of 800-by-600 and a vertical refresh rate of 72 Hz.  Stimuli consisted of oblique sinusoidal gratings windowed by a $10$ degree square centered vertically to the left or right of fixation, depending on the hemifield being stimulated.   Stimuli for each hemified were mirror symmetric, with gratings on the left oriented at $45\,^{\circ}$, and those on the right oriented at $135\,^{\circ}$.  For both hemifields, the spatial frequency of the gratings was $3$ cycles per degree, with mean luminance kept constant throughout the experiments.  Stimulus contrast was defined as the difference between the maximum and minimum luminance of the grating divided by their sum. The contrast of the stimulus was temporally modulated (i.e., contrast reversal) by a 9 Hz sinusoid. Each stimulus presentation consisted of ten 1-second presentations of contrast reversal. Each 1s presentation occurred at a fixed contrast, with the first set to 0.05, the last at 0.8, and the rest logarithmically spaced between these two values. The 22 subjects were split into two groups of 11. For one group, the stimulus was presented 90 times in the right visual field, and 10 times in the left visual field, with the ordering randomized before the beginning of the session. For the other group, these numbers were reversed.  In the analysis, we retained only the 90 trials corresponding to the predominantly stimulated hemifield.

The EEG was acquired using a $128$-channel electrode array (Electrical Geodesics Inc, OR) at a sampling rate of $500$Hz with a vertex reference electrode.  All pre-processing was done offline using in-house software.  Signals were band-pass filtered between $0.1$ Hz and $200$ Hz.  Channels in which $15\%$ of the samples exceeded a fixed threshold of $2550$  $\mu V$ were replaced with a spatial average of the six nearest neighbors.  Within each channel, 1 second epochs containing samples exceeding a fixed threshold ($2550$ $\mu V$) were rejected.   The EEG was then re-referenced to the common average of all channels. Spectral analysis was performed via a Discrete Fourier Transform with 0.5 Hz resolution. The contrast reversing stimuli generated VEPs whose spectra were dominated by even multiples of the presentation frequency (i.e., 2nd, 4th and 6th harmonics); as such, the real- and imaginary-components of these 3 Fourier coefficients across the array formed the $128$-by-$6$ data record stemming from each trial.

\section*{Results}
\noindent To evaluate the proposed technique, we applied RCA to two SSVEP data sets whose full details are described in the Methods.  Briefly, the first is a synthetic data set which employs a BEM head model to simulate the propagation of cortical signals to an array of scalp electrodes; the simulation analysis allows for ground-truth measurements of the SNR as well as a comparison of recovered component topographies with the lead fields of the underlying sources.   The second (real) data set was acquired in a paradigm consisting of visual stimulation of the left- or right-hemifield with sinusoidal gratings presented at a temporal frequency of $f=9$Hz.    In addition to RCA, we evaluated the popular CSP \cite{blankertz2008optimizing} method as well as PCA.  
\begin{figure*}[htb]
\centering
\includegraphics[width=150mm]{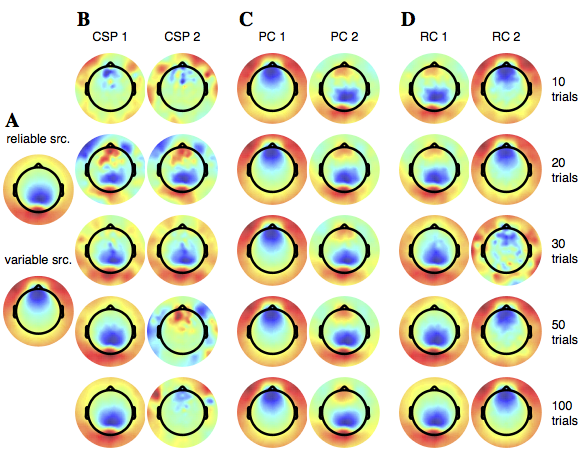}
\caption{\label{fig:simComps} Sample component topographies recovered by component analysis techniques as a function of the number of available trials per simulation draw. ({\textbf{A}}) Lead fields of the reliable (top panel) and variable (bottom panel) sources employed in the simulation, as computed from a three-layer BEM head model.  ({\textbf{B}}) CSP requires 30 trials to recover the topography of the desired source, while unable to cleanly extract the lead field of the variable source, even at 100 trials.  ({\textbf{C}}) For all number of trials, PCA recovers the variable source (PC1) and desired source (PC2).  However, due to PCA's requirement of spatially orthogonal weight vectors, the topography of PC2 contains prominent contributions from the frontal interference source, even at 100 trials.  ({\textbf{D}}) RCA recovers the reliable source (RC1) and variable source (RC2), with clean versions of both found at 50 trials. }
\end{figure*}

We first present the results of evaluating the three component analysis techniques on the synthetic data set, beginning with an examination of the extracted components.    When comparing the scalp topographies of the various components, we depict not the weights themselves (i.e., the ``${\bf{W}}$'') but rather their projection onto the scalp (i.e., the ``${\bf{A}}$'', see Methods).  For a detailed explication of the computation of this scalp projection, please refer to the Methods section ``Component scalp projections'' and \cite{haufe2014on,parra2005recipes}.

Figure \ref{fig:simComps} illustrates the scalp projections of the first two components recovered by each method, where we have chosen a representative draw from the Monte Carlo simulation to construct the figure.  The lead fields corresponding to the ground-truth signal sources are illustrated in Panel A: the ``reliable'' source has a symmetric front-to-back dipolar topography roughly centered over electrode Oz, and serves as the desired signal in the simulation.  Meanwhile, the ``variable'' source is marked by a symmetric frontal topography with the center of the negative pole roughly over electrode Fz; this source serves as the interference.  Panel B depicts the topographies yielded by the CSP technique.  Given 10 trials of data, the technique is unable to recover a component with an occipital topography resembling the desired source.  This topography begins to emerge at 20 trials (CSP1 and CPS2), though it is visibly noisy and contains strong contribution from the frontal interference source.  By increasing the available data from which to learn the CSP spatial filters, the ground-truth lead field emerges cleanly at 50 trials.  Moreover,  a noisy version of the variable source is also recovered in CSP2.  

Consider next Figure \ref{fig:simComps}C, which depicts the scalp projections yielded by PCA.  For all values of the number of available trials, the first PC is clearly capturing the frontal interference source.  As this source exhibits more variability  than the desired source (i.e., amplitude and phase differ across trials), the projection of the data which maximizes the proportion of variance explained is matched to its frontal scalp topography.  Meanwhile, for all values of the number of trials, the desired signal is captured in PC2.  Note, however, that the resulting component is clearly a mixture of the desired and interference sources, as evident by its frontal activity.  This mixture ensues as a result of PCA's constraint of spatial orthogonality among its weight vectors.  

Figure \ref{fig:simComps}D illustrates the component projections yielded by RCA: at 10 trials per draw, the method is able to recover both desired (RC1) and interfering (RC2) sources. The topography of RC1 ``loses'' its frontal positivity at 30 trials, and yields clean topographies for both simulated sources at 50 trials.  At 100 available trials, RCA cleanly recovers both reliable and variable sources from the data.  The correct ordering of recovered components stems from the technique's criterion of reliability: the desired source is, by construction, more reliable across trials than the variable source.   Moreover, the ``cleanliness'' of the recovered topographies is aided by its explicit use of the trial dimension, which focuses the spatial filter weights on dimensions which maximally covary across these trials.  

\begin{figure*}[htb]
\centering
\includegraphics[width=150mm]{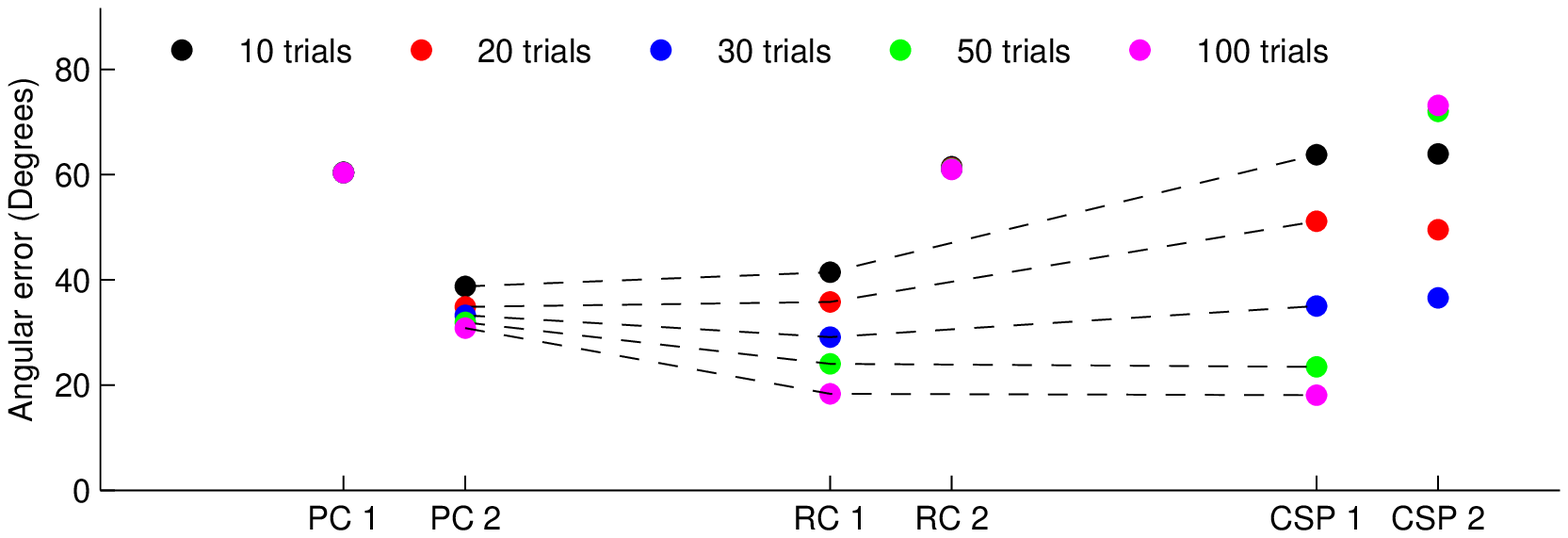}
\caption{\label{fig:simSRC} Assessing the ability of component analysis techniques to recover the desired source in a synthetic data set.  Each data point conveys the median (across Monte Carlo simulation draws) vector angle between the recovered component's scalp projection and the lead field of the (known) desired signal source.  The dashed line connects the component best matched to the underlying lead field for each method.  At 10-20 trials, the topographies of PC2 and RC1 are closest to the desired lead field, with the yielded angular errors significantly lower than that of CSP1 ($p<3 \times 10^{-14}$, paired left-tailed Wilcoxon rank sign test).   At 30 trials, RC1 recovers the component closest to the generating lead field ($p=0.04$).   At 50-100 trials, CSP1 slightly outperforms RC1, which in turn yields lower median angular error than PC2 ($p<6 \times 10^{-7}$).}
\end{figure*}

Figure \ref{fig:simComps} was derived from a representative but single simulation draw; a more complete evaluation of the recovered components is shown in Figure \ref{fig:simSRC}, which depicts the median deviations (across simulation draws) between the obtained component scalp projections and the reliable signal's lead field.  That is, we use the vector angle between the obtained and desired topography as a measure of ``goodness''.   The dashed line connects the components with the lowest angular deviation for each method (i.e., PC2, RC1, and CSP1).  At 10 trials per draw (black markers), PC2 and RC1 yield the lowest median angular error ($39^o$ and $41^o$ respectively, no significant difference); these median values are significantly lower than that of CSP1 (median=64$^o$, $n=500$ $p=2\times10^{-29}$, paired, left-tailed Wilcoxon signed rank test).    At 20 trials per draw, the median deviations are reduced to $35^o$ for PC2, $36^o$ for RC1, and $51^o$ for CSP1.  PC2 and RC1 are found to significantly outperform CSP1 ($p<3 \times 10^{-14}$).  At 30 trials per draw, RC1 yields the lowest median error of $29^o$, followed by $33^o$ for PC2, and $35^o$ for CSP1 (all pairwise differences are significant, $p<0.04$).   At 50 trials per draw, CSP1 yields the lowest angular deviation of $23^o$, followed by RC1 at $24^o$, and PC2 at $32^o$.  All pairwise differences are again statistically significant ($p<6\times10^{-7}$).   Finally, given 100 trials per draw, CSP1 yields an angular error of $18.0^o$, with the corresponding errors for RC1 and PC1 being $18.3^o$ and $31^o$, respectively, with all pairwise differences significant ($p<2\times10^{-35}$).   

\begin{figure*}[htb]
\centering
\includegraphics[width=150mm]{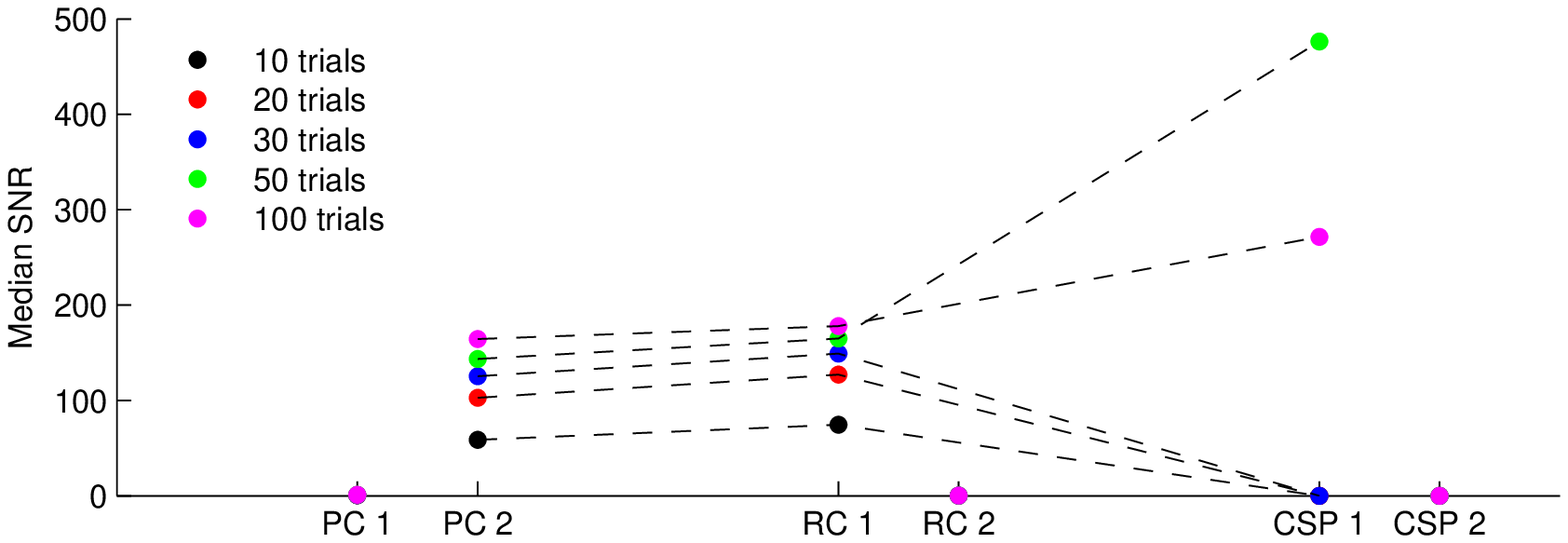}
\caption{\label{fig:simSNR}  A comparison of the SNRs yielded by component analysis techniques on a synthetic data set, shown as a function of the number of trials.  The data points convey the median value across simulation draws.  The dashed line connects the components with the highest SNR for each method.    At $10$ trials, the highest SNR is attained by RC1 (median=74), followed by PC2 (median=58) and then CSP1 (median=0).   This ordering is preserved up to 30 trials and all pairwise differences are statistically significant ($p<1\times 10^{-32}$,  paired, right-sided Wilcoxon signed rank test).  At 50-100 trials, however, CSP1 yields the highest SNR (median = 271 at 100 trials), followed by RC1 (median = 178) and PC2 median = 164).  All pairwise differences are statistically significant ($p<3 \times 10^{-28}$).  The behavior of CSP on this data set is a reflection of the technique's aim to project the data onto the null space of the noise, which it is here able to find at 50 trials. } 
\end{figure*}

While the physiological plausibility of an extracted component is certainly important to inferring the corresponding source, in some applications (for example, signal detection), it may be appropriate to sacrifice physiological meaning in order to achieve a high SNR.  This entails focusing the spatial filter weights on the channels which exhibit low noise power.  To that end, 
Figure \ref{fig:simSNR} displays the SNRs yielded by the components found above as a function of the number of available trials.  The data points convey the median value across simulation draws, while the dashed line connects the components with the highest SNR for each method.    At $10$ trials, the highest SNR is attained by RC1 (median=74), followed by PC2 (median=58) and then CSP1 (median=0).   This ordering is preserved up until 30 trials, with all pairwise differences showing statistical significance ($p<1\times 10^{-32}$,  $n=500$, paired right-sided Wilcoxon rank sign test).  At 50-100 trials, however, CSP1 yields the highest SNR (median = 271 at 100 trials), followed by RC1 (median = 178) and PC2 (median = 164).  All pairwise differences are again statistically significant ($p<3 \times 10^{-28}$).   The behavior of CSP on this data set is a reflection of the technique's aim to project the data onto the null space of the noise, which it is here able to find at 50 trials. 

\subsection*{Evaluation on real data.}
We now turn to the evaluation of the component analysis techniques on a real data set.  Similar to what was described above for the simulated data, we sought to evaluate the physiological plausibility of the components yielded by RCA and its alternatives.  Here, however, we do not possess ground-truth information as to the lead fields of the underlying cortical sources.   From the anatomy of the human visual system, however, input in the left visual field (i.e., left ``hemifield'') is processed in the right cerebral hemisphere and vice versa.   Thus, one way of assessing the physiological relevance of the obtained components is to compute the projections separately for stimulation of the left and right hemifields (LH and RH, respectively), and then observe whether a contralateralization of the scalp topographies emerges.  Moreover, to assess the role of the input SNR in the physiological plausibility of the resulting components, we performed the analysis separately for varying levels of stimulus contrast and thus response amplitude.  In what follows, we focus exclusively on the first component (i.e., CSP1, PC1, and RC1) of each candidate method.  

\begin{figure*}[htb]
\centering
\includegraphics[width=150mm]{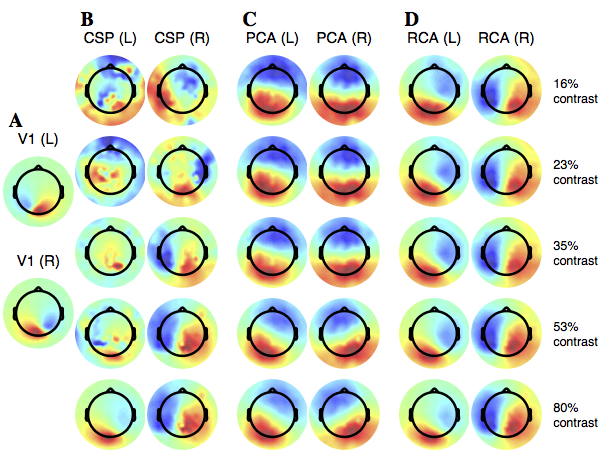}
\caption{\label{fig:alexComps} RCA yields physiologically plausible scalp topologies insensitive to input SNR.   ({\bf{A}})  Theoretical scalp projections from the left (L) and right (R) primary visual cortices as computed by the boundary element model (BEM) of an individual human head.   {\bf{(B)}} The scalp projections of the first CSP as computed from data recorded during visual stimulation of the left (L) and right (R) hemifields.  At low contrasts, the topologies lack physiological plausibility and appear to to be driven by noise; moreover,  a lack of lateralization is apparent until the highest contrast level.  {\bf{(C)}} Same as (A) but now for the first principal component (PC).  The corresponding topographies bear closer resemblance to the dipolar maps commonly seen in EEG forward modeling. The level of observed contralateralization increases with the stimulus contrast, highlighting that PCA components become more physiologically plausible as the input SNR increases.    ({\bf{D}}) The scalp projections of the first reliable component (RC) exhibit physiologically plausible topographies with clear lateralization even at low contrast-values, with the topographies remaining relatively stable over the entire contrast range.  This suggests that the technique is more robust to a low input SNR, which follows from its exploitation of the trial-to-trial covariance structure in the data.   Moreover, the topographies bear a close resemblance to the lead fields from primary visual cortex (Panel A), including the location of the back end of the dipole.  Note that the broader poles of the empirical topographies are likely due to the smoothing out of the projection by averaging across multiple subjects.    } 
\end{figure*}

Figure \ref{fig:alexComps}A displays the lead fields from both left and right primary visual cortices (V1-L and V1-R, respectively), as computed from a BEM model of a sample head.  Striate visual cortex is expected to be a major generator of the activity evoked by this SSVEP paradigm.  The dipolar topographies exhibit mirror symmetry with the left primary visual cortex projecting positively to the right occipital electrodes, and vice versa.  

Figure \ref{fig:alexComps}B depicts the scalp projections of CSP for both LH and RH at each stimulus contrast.  At low contrast, the topographies are visibly noisy and lack the spatial structure expected in a visual paradigm (i.e., concentration of activity at the occipital electrodes).  A contralateralization of the scalp topographies with stimulated hemifield is not apparent until 80\% contrast, at which the scalp projections still lack strong mirror-symmetry.  Consider now Panel C, which displays the topographies of PCA.  The ensuing dipolar topographies more closely resemble the maps expected from this visual paradigm.  Moreover, there is a progressively greater level of mirror symmetry in the obtained scalp projections with increasing contrast, and clear contralateralization emerges at 53\% contrast.  The scalp projections of RCA are shown in Panel D: a contralateralization with the stimulated hemifield is readily observed at 16\% contrast.   Moreover, the topographies remain quite stable with contrast, suggesting that RCA is robust to input SNR.    Finally, note that the RCA topographies bear the closest resemblance to the V1 lead fields of Panel A.  We refrain, however, from proclaiming the RCs as better recovering the underlying sources in this data set: extrastriate visual areas such as V2, V3, V3a, V4, and MT, whose lead fields also contralateralize with the stimulated hemifield, may also have been activated.   


\begin{figure*}[htb]
\centering
\includegraphics[width=150mm]{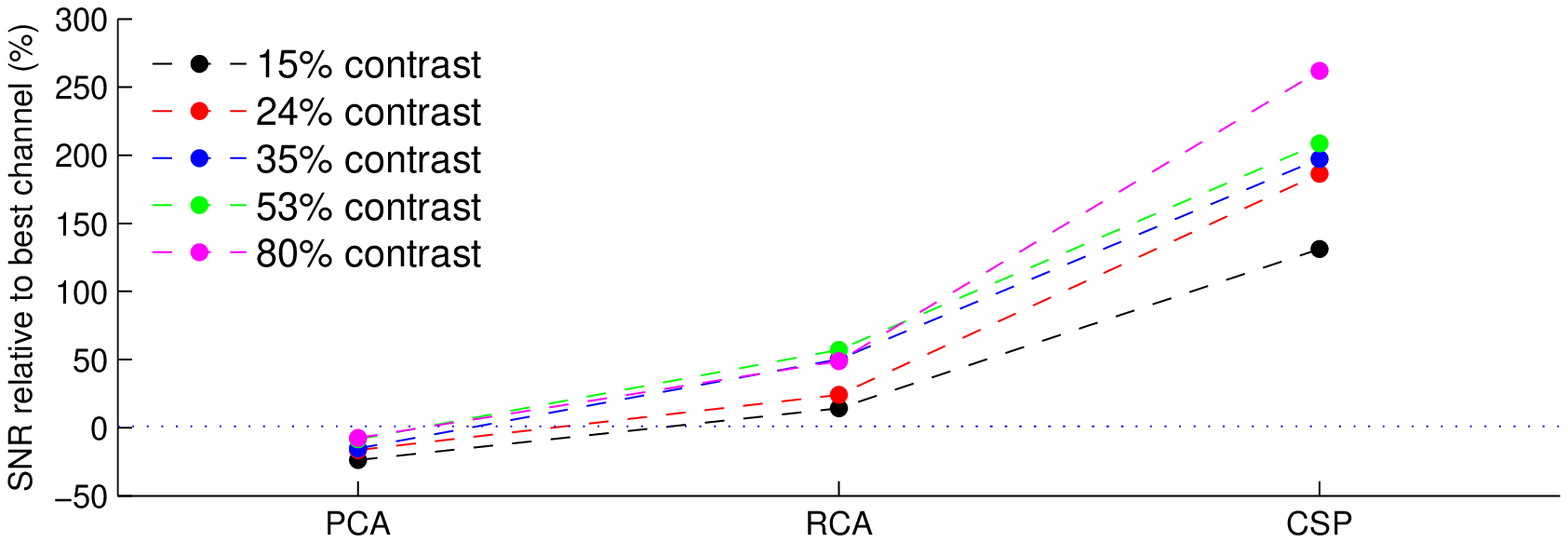}
\caption{\label{fig:alexSnr} RCA yields SNRs greater than the best single electrode.  The figure shows the median (across $n=1980$ trials pooled across subjects) SNR improvement over the best (post-hoc selected) individual electrode as yielded by various component analysis methods.  For all contrast levels, PCA yields degradations in SNR relative to the best channel.  RCA provides a $14\pm7$\% improvement in the low-contrast condition, while yielding $49\pm18$\% improvement in the high-contrast condition.  For all contrast values, the median RCA SNR is significantly greater than that of PCA ($p<3\times10^{-22}$, paired, right-tailed Wilcoxon signed rank test).   Meanwhile, the CSP method explicitly optimizes the SNR and offers improvements as high as $262\pm52$\% in the high-contrast case.   The median CSP SNR is significantly higher than the corresponding median SNR yielded by RCA ($p<6\times10^{-69}$) for all stimulus contrasts.} 
\end{figure*}

Next, we computed the single-trial SNR of the components found by the three candidate methods, computing the spatial filter weights individually for each subject to take into account inter-subject variability.   SNR estimation was facilitated by defining noise frequency bands as lying directly adjacent to the frequencies of interest (i.e., the first three even harmonics).   We pool across the subject dimension to yield a distribution of $n=1980$ single-trial SNRs for each method and stimulus contrast level.  Additionally, we performed a post-hoc exhaustive search of the electrode space to identify the single electrode yielding the highest single-channel SNR.    The results are shown in Figure \ref{fig:alexSnr}, where the data points depict the median SNR improvement over this best individual-channel.    

Notice first that the median SNR yielded by the PCA is lower than that given by the best individual channel for all contrast values: at low (16\%) contrast, PCA suffers a median SNR degradation of $24\pm7$\%  (mean $\pm$ s.e.m.) relative to the best individual electrode.  This is indicative of the fact that dimensions explaining the majority of the variance in EEG often capture noise sources.   At high (80\%) contrast, the degradation is lower:  $8\pm12$\%. Meanwhile, RCA offers a median SNR improvement of $14\pm7$\% at low-contrast, and $49\pm18$\% at high contrast, relative to the best channel.  Finally, CSP explicitly optimizes the SNR and, coupled with the large number of trials per subject,  yields large improvements over the best channel: $131\pm116$\% at low-contrast, and $262\pm52$\% at high contrast.  We performed a Wilcoxon signed rank test to determine whether the differences in SNR improvements between methods are significant:  for all input contrasts, the SNRs yielded by RCA are significantly greater from those yielded by PCA ($p<3\times10^{-22}$, $n=1980$, paired, right-tailed Wilcoxon signed rank test).    Similarly, the CSP SNRs are significantly greater than those of RCA ($p<6\times10^{-69}$).  

\begin{figure*}[htb]
\centering
\includegraphics[width=150mm]{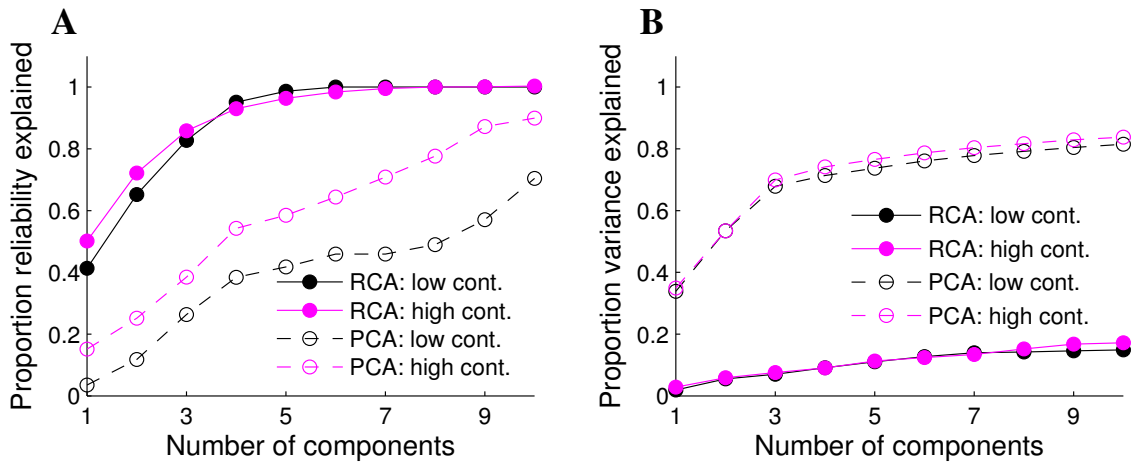}
\caption{\label{fig:propRel} Reducing dimensionality of SSVEP data sets by forming components optimizing the reliability and variance. {\bf{(A)}}  Proportion of reliability explained as a function of the number of retained components.  The first four RCs capture $>95$ of the trial-to-trial covariance in the data, with $35-55\%$ captured by the first four PCs {\bf{(B)}} Proportion of within-trial variance explained as a function of the number of retained components. The first four PCs explain $75\%$ of the variance, with the first four RCs capturing $<10\%$ of the within-trial variance. } 
\end{figure*}
To quantify the level of dimensionality reduction afforded by RCA on this data set, we computed the \emph{proportion of reliability explained} as a function of the number of RCs (see equation \ref{eqn:propRelExpl} in Methods).   This is the reliability analogue of the proportion of variance explained as a function of the number of PCs, which was also computed on the data.  To shed light on the tradeoff between trial-to-trial covariance and variance explained, we then performed a ``cross-over'' analysis which considers the amount of trial-to-trial reliability captured by the PCs, and the amount of variance explained by the RCs.  

Figure \ref{fig:propRel}A displays the proportion of reliability explained as a function of the number of retained RCs (solid line, filled markers) and PCs (dashed line, open markers) for the low (black markers) and high (magenta markers) contrast case.  The first RC captures 41\% and 50\% of the trial-to-trial covariance for the low- and high-contrast case, respectively.  Meanwhile, more than $93\%$ of the reliability is contained in the first four RCs, representing a dimensionality reduction of $128/4=32$ while only sacrificing less than 10\% of the reliable SSVEP.  Meanwhile, the PCs are tuned to optimally capture within-trial variance and thus capture substantially less of the trial-to-trial reliability: the first four PCs capture 38\% and 54\% of the trial-to-trial covariance for the low- and high-contrast data, respectively.  Moreover, while the cumulative curve converges to unity at $C=6$ RCs, more than ten PCs are required to explain the full reliability spectrum.     

Figure \ref{fig:propRel}B displays the corresponding amount of variance explained as a function of the number of retained PCs and RCs.   The first four PCs explain $71\%$ and $74\%$ of the within-trial variance for the low- and high-contrast case, respectively.   Note, however, that the number of PCs required to account for the bulk (i.e., virtually all) of the within-trial variance is still substantially greater than ten.  Meanwhile, the first four RCs capture just 9\% of the within-trial variance, exemplifying the stark difference in criteria being optimized by PCA and RCA.

\section*{Discussion}

\noindent We have presented a novel component analysis technique which drastically reduces the dimensionality of SSVEP data sets while retrieving physiologically plausible scalp topographies and yielding SNRs greater than the best single electrode.   The method follows from the fundamental assumption of evoked responses, namely that the neural activity evoked by the experimental paradigm is reproducible from trial-to-trial.  

The results of the simulation study revealed that RCA provides a desirable tradeoff between physiological plausibility and output SNR.  Specifically, the technique recovered the desired source topography with comparable or lower angular error relative to PCA and CSP for all number of trials per simulation draw (Figures \ref{fig:simComps} and \ref{fig:simSRC}).  Moreover, RCA achieves the highest output SNR at $\leq$ 30 trials, with CSP significantly outperforming RCA/PCA at $\geq$ 50 trials (Figure \ref{fig:simSNR}).  
 
The analysis of real data acquired in a single-hemifield visual stimulation paradigm demonstrated the ability of RCA to yield physiologically plausible components which contralateralize with the stimulated hemifield even at low contrast (Figure \ref{fig:alexComps}).    PCA recovered components which exhibit progressively more physiological plausibility with increasing contrast; however, the output SNR of these extracted components fell below that yielded by the best single channel.  Finally, CSP yielded high SNR components for all contrast levels, but sacrificed physiological plausibility of the topographies, particularly at low contrast.    

\emph{Relevance of simulation results.}  The simulation study performed here is clearly a simplification of the actual neural environment that generates SSVEPs.  Aspects of the simulation that reflect potential deviations from reality include: dipolar sources \cite{riera2012pitfalls},  average isotropic conductivity values \cite{wolters2006influence}, and uncorrelated additive sensor noise.   Nevertheless, we feel that simulations are useful here in order to quantify the tradeoffs inherent to the existing methods.  While the figures of merit obtained in a simulation may not necessarily translate to real settings, their relative values (i.e., comparisons across methods) are more likely to hold.  Moreover, the validity of simulation results is boosted when consistent with that found in evaluations on real data.  Notice, for example, the general agreement between the results of Figures \ref{fig:simComps} and \ref{fig:alexComps}.

\emph{Learning on individual versus aggregated data.}  As with any component analysis technique, a learning procedure is employed by RCA to estimate the reliability-maximizing spatial filters.  An important question is whether one should learn on subject-aggregated data, yielding a set of uniform RCs for the entire data set, or rather compute the components separately for each subject.  The tradeoff here is between the noise level in the estimated covariance matrices (subject-aggregated covariance has lower estimation noise) and the ability to exploit individual differences in component topographies.  For example, to construct Figure \ref{fig:alexComps}, which aimed to characterize the reliable activity evoked by the visual stimulation paradigm, we pooled data from all subjects to learn the (smoothed) RCA spatial filters.   On the other hand, when comparing achieved SNR across component analysis methods in Figure \ref{fig:alexSnr}, we opted instead to learn the optimal filters individually for each subject, as structural and functional variations are expected to lead to disparate topographies.  Note that when learning the filter weights individually, the resulting dimensionality-reduced data is not congruent across subjects (i.e., RC1 of subject 1 generally lies in a different space than RC1 of subject 2).  As such, care should be taken when comparing the projected spectra across data computed using different spatial filters.   

\emph{RCA as a reliability filter.}  Here we have focused the analysis to the space of the formed components which encompass data integrated across multiple electrodes.  In some cases, it may be desirable to rather analyze the data in the original electrode space.  To that end, it is also possible to treat RCA as a ``reliability filter'' which outputs a data set whose dimensionality is that of the original data set.  This is achieved by first projecting the data onto a set of $C$ RCs, and then back-projecting this rank-reduced data onto the scalp: if $\bf{Y}$ denotes the frequency-by-component RCA data matrix, and the electrode-by-component matrix $\bf{A}$ denotes the corresponding scalp projections (see Methods), then the reconstructed sensor-space data matrix follows as $\tilde{{\bf{X}}}= {\bf{A}} {\bf{Y}}^T $ .  This procedure removes dimensions exhibiting low trial-to-trial reliability, presumably corresponding to noise sources, from the data.  Conventional analysis methods such as trial-averaging may then be employed on $\tilde{{\bf{X}}}$.  

\emph{Application to source imaging.}  We have shown here that the topographies of the various RCs bear strong resemblance to the underlying lead fields generating the observed SSVEP.  This suggests that RCA may be combined with source localization approaches to yield robust estimates of the location of the neuronal generators.  Note that the conventional manner of performing EEG source localization is to select an array of scalp potentials at a given latency (or frequency-band) and project the resulting vector onto the cortical surface using an appropriately generated inverse matrix.  While we do not engage here in a discussion of the legitimacy of the resulting source location estimates, we do propose that the RCA scalp projections (for example, Figure \ref{fig:alexComps}D) may themselves be employed as inputs into a source localization algorithm.   Note that these scalp projections are not tied any particular time instant: rather, they correspond to the source of activity which is reliably evoked across trials.  As such, their use as source localization inputs eliminates the need to choose a particular time instant at which to localize.  The resulting cortical source distribution (i.e., the output of the source localization) bears a time course given by the RC whose scalp projection was used for the input.  Moreover, this procedure yields insight into the sources underlying the reliable activity extracted by the component, and whether the RCs represent vast mixtures of generators or more spatially localized dipoles.

\emph{Application to BCIs.}  The proposed technique is primarily aimed at data sets collected in a neurobiological imaging setting.  However, we anticipate that RCA may also become relevant for BCI applications which learn patterns of electrodes associated with a particular cognitive state.  We propose that RCA be employed at the front-end as a feature selection step which reduces the dimensionality of the input feature vector while still capturing the reproducible neural features.  Note that the technique is inherently blind, requiring only multiple congruent data records (i.e., without labels indicating the outcome of any associated task) to learn the RCs.   Moreover, the resulting features may yield better generalization due to their closer link to physiology.  

\emph{Frequency- vs time-domain}.   We note that many spatial filtering approaches to SSVEPs employ a time-domain formulation.  Here, we advocate a frequency-domain approach which replaces temporal samples of the electric potential with Fourier coefficients at the presentation frequency and/or its associated harmonics.  Moreover, we only retain the dominant response frequencies (i.e., spikes in the Fourier spectrum) in the feature vector inputted to RCA.  This provides relative immunity from artifacts and a minimally redundant signal.   Note that the trial-aggregation procedure of RCA (see equation \ref{eqn:agg}) effectively increases the column dimensionality of the data records (i.e., by aggregating all pairs of trials, one has effectively reduced the number of trials required to obtain a full-rank sample covariance matrix).   Nevertheless, it is worthwhile to mention that RCA is also applicable to time-domain data (i.e., the second dimension of ${\bf{X}}_n$ being time, not frequency).  This applies to transient evoked potential studies in which it is more appropriate to represent the neural responses as time series.  In this version, the spatial filters seek to maximize the \emph{temporal correlation} between the activity evoked during each trial.  As the dimensionality of the temporal dimension is expected to be significantly larger than that of the spectral dimension of SSVEPs, a larger number of trials is typically required in order to recover the underlying signal sources.  

\emph{Goodness of the SNR as a quality metric}.  The SNR is a natural metric which certainly conveys the most obvious goal of a signal processing algorithm: reducing the noise.  However, it is worthwhile to point out that SNR becomes infinite for zero-noise even if the desired signal has been greatly distorted.  In other words, the SNR down-weights signal distortion in favor of noise reduction.   However, electrodes which possess the lowest noise levels are not necessarily those at which the cortical sources project to most strongly.  Thus, we caution from interpreting the SNR as a ``gold-standard'' in measuring the goodness of a spatial filtering algorithm.  There are cases in which one may be willing to sacrifice noise reduction in order to obtain a minimally distorted version of the underlying source signal.  To that end, several approaches to managing the tradeoff between  signal distortion and noise reduction have been proposed in related signal processing fields \cite{chen2006new}.

\emph{Emergence of reliability in neuroscience.}  Our findings add to the emerging body of evidence pointing to the utility of employing reliability as a criterion with which to measure and extract meaningful neural signals.   Highly reliable neural responses have been observed in extensive parts of cortex during naturalistic audio(visual) stimulation in fMRI \cite{hasson2004intersubject}, EEG \cite{dmochowski2012correlated,dmochowski2014audience}, and magnetoencephalography (MEG) \cite{koskinen2014uncovering,lankinen2014intersubject}.   Moreover, a recent fMRI study has reported that the level of inter-subject correlation in the blood-oxygenation-level-dependent (BOLD) signal is greater when the stimulus is presented in 3D \cite{gaeblerlabel2014stereoscopic}.  In terms of trial-based applications, a method to identify correlations among spectral envelopes of multivariate electrophysiological recordings has been proposed in \cite{dahne2014finding}.   Finally, reproducibility of neural activation has been linked to conscious perception \cite{schurger2010reproducibility}.   Collectively, these findings highlight the increasing use of reliability as a meaningful feature in neuroscience: indeed, data collection in the brain sciences almost always encompasses multiple data records (i.e., multiple trials, multiple subjects, or both).  Given that the desired signal is expected to be common to these records, reliability represents a natural means of separating the reliable signal from the variable noise.  

\emph{Application matters.}  While we have presented RCA as an alternative to commonly employed methods such as CSP and PCA, we do not suggest that it is the ``best'' component analysis method for analyzing SSVEPs.  Rather, we feel that the field of cognitive neuroimaging has not reaped the benefits of spatial filtering approaches in the same way that the BCI world has.  For BCIs, SNR may in fact be the most appropriate metric, as it may best relate to information bit rate.  However, for elucidating neural processing in human visual cortex, extracting components which have physiological relevance is of utmost importance.  Here, we believe that exploiting the trial-to-trial reliability of evoked responses is an appropriate way of bringing the recovered components closer to physiology.




\section*{References}

\begin{thebibliography}{31}
\providecommand{\natexlab}[1]{#1}
\providecommand{\url}[1]{\texttt{#1}}
\expandafter\ifx\csname urlstyle\endcsname\relax
  \providecommand{\doi}[1]{doi: #1}\else
  \providecommand{\doi}{doi: \begingroup \urlstyle{rm}\Url}\fi

\bibitem[Ales et~al.(2010)Ales, Yates, and Norcia]{ales2010v1}
J.~M. Ales, J.~L. Yates, and A.~M. Norcia.
\newblock V1 is not uniquely identified by polarity reversals of responses to
  upper and lower visual field stimuli.
\newblock \emph{Neuroimage}, 52\penalty0 (4):\penalty0 1401--1409, 2010.

\bibitem{almoqbel2012technique}
F. Almoqbel, S. J. Leat, and E. Irving
\newblock The technique, validity and clinical use of the sweep VEP
\newblock \emph{Ophthalmic Physiol Opt.} 28(5):393-403. 2008.

\bibitem{andersen2012bottom}
S. K. Andersen, M. M. Muller, and J. Martinovic. 
\newblock Bottom-up biases in feature-selective attention.
\newblock \emph{The Journal of Neuroscience} 32.47: 16953-16958. 2012). 

\bibitem{bin2009online}
G. Bin, X. Gao, Z. Yan, B. Hong, and S. Gao, S. (2009).
\newblock An online multi-channel SSVEP-based brain-computer interface using a canonical correlation analysis method. 
\newblock \emph{Journal of neural engineering}, 6(4), 046002, 2009.

\bibitem[Blankertz et~al.(2008)Blankertz, Tomioka, Lemm, Kawanabe, and
  Muller]{blankertz2008optimizing}
B.~Blankertz, R.~Tomioka, S.~Lemm, M.~Kawanabe, and K.-R. Muller.
\newblock Optimizing spatial filters for robust eeg single-trial analysis.
\newblock \emph{Signal Processing Magazine, IEEE}, 25\penalty0 (1):\penalty0
  41--56, 2008.
 
\bibitem{chen2006new}
J. Chen, J. Benesty, Y. Huang, and S. Doclo, S. 
\newblock New insights into the noise reduction Wiener filter. 
\newblock \emph{IEEE Transactions on Audio, Speech, and Language Processing},14(4), 1218-1234. 2006
  
\bibitem{dahne2014finding}  
S. Dahne, V. V. Nikulin, D.  Ramírez, P. J. Schreier, K. R. Muller, K. and S. Haufe, S. 
\newblock Finding brain oscillations with power dependencies in neuroimaging data
\newblock \emph{NeuroImage}, 96, 334-348. 2014. 
  

\bibitem[Dmochowski et~al.(2012)Dmochowski, Sajda, Dias, and
  Parra]{dmochowski2012correlated}
J.~P. Dmochowski, P.~Sajda, J.~Dias, and L.~C. Parra.
\newblock Correlated components of ongoing eeg point to emotionally laden
  attention--a possible marker of engagement?
\newblock \emph{Frontiers in human neuroscience}, 6, 2012.

\bibitem{dmochowski2014audience}
J.~P. Dmochowski, M. A. Bezdek, B. P. Abelson, J. S. Johnson, E. H. Schumacher and L.~C. Parra.
Audience preferences are predicted by reliability of temporal neural processing.
 \emph{Nature Communications}, 2014.


\bibitem[Friman et~al.(2007)Friman, Volosyak, and Graser]{friman2007multiple}
O.~Friman, I.~Volosyak, and A.~Graser.
\newblock Multiple channel detection of steady-state visual evoked potentials
  for brain-computer interfaces.
\newblock \emph{Biomedical Engineering, IEEE Transactions on}, 54\penalty0
  (4):\penalty0 742--750, 2007.

\bibitem{gaeblerlabel2014stereoscopic}
M. Gaeblerlabel, F. Biessmannlabel, J. P. Lamke, K. R. Muller, H. Walter, and S. Hetzer, S. 
\newblock Stereoscopic depth increases intersubject correlations of brain networks. 
\newblock \emph{NeuroImage}, 2014.

\bibitem[Garcia-Molina and Zhu(2011)]{garcia2011optimal}
G.~Garcia-Molina and D.~Zhu.
\newblock Optimal spatial filtering for the steady state visual evoked
  potential: Bci application.
\newblock In \emph{Neural Engineering (NER), 2011 5th International IEEE/EMBS
  Conference on}, pages 156--160. IEEE, 2011.

\bibitem[Goldenholz et~al.(2009)Goldenholz, Ahlfors, H{\"a}m{\"a}l{\"a}inen,
  Sharon, Ishitobi, Vaina, and Stufflebeam]{goldenholz2009mapping}
D.~M. Goldenholz, S.~P. Ahlfors, M.~S. H{\"a}m{\"a}l{\"a}inen, D.~Sharon,
  M.~Ishitobi, L.~M. Vaina, and S.~M. Stufflebeam.
\newblock Mapping the signal-to-noise-ratios of cortical sources in
  magnetoencephalography and electroencephalography.
\newblock \emph{Human brain mapping}, 30\penalty0 (4):\penalty0 1077--1086,
  2009.

\bibitem[Golub and Van~Loan(2012)]{golub2012matrix}
G.~H. Golub and C.~F. Van~Loan.
\newblock \emph{Matrix computations}, volume~3.
\newblock JHU Press, 2012.

\bibitem[Hallez et~al.(2007)Hallez, Vanrumste, Grech, Muscat, De~Clercq,
  Vergult, D'Asseler, Camilleri, Fabri, Van~Huffel, et~al.]{hallez2007review}
H.~Hallez, B.~Vanrumste, R.~Grech, J.~Muscat, W.~De~Clercq, A.~Vergult,
  Y.~D'Asseler, K.~P. Camilleri, S.~G. Fabri, S.~Van~Huffel, et~al.
\newblock Review on solving the forward problem in eeg source analysis.
\newblock \emph{Journal of neuroengineering and rehabilitation}, 4\penalty0
  (1):\penalty0 46, 2007.

\bibitem[Hasson et~al.(2004)Hasson, Nir, Levy, Fuhrmann, and
  Malach]{hasson2004intersubject}
U.~Hasson, Y.~Nir, I.~Levy, G.~Fuhrmann, and R.~Malach.
\newblock Intersubject synchronization of cortical activity during natural
  vision.
\newblock \emph{science}, 303\penalty0 (5664):\penalty0 1634--1640, 2004.

\bibitem{haufe2014on}
S. Haufe, F. Meinecke, K. Gorgen, S. Dahne, J. D. Haynes, B. Blankertz, F. Biessmann.  
\newblock On the interpretation of weight vectors of linear models in multivariate neuroimaging. 
\newblock \emph{NeuroImage}, 87, 96-110.  2014.

\bibitem[Herrmann(2001)]{herrmann2001human}
C.~S. Herrmann.
\newblock Human eeg responses to 1--100 hz flicker: resonance phenomena in
  visual cortex and their potential correlation to cognitive phenomena.
\newblock \emph{Experimental brain research}, 137\penalty0 (3-4):\penalty0
  346--353, 2001.

\bibitem[Hotelling(1936)]{hotelling1936relations}
H.~Hotelling.
\newblock Relations between two sets of variates.
\newblock \emph{Biometrika}, pages 321--377, 1936.

\bibitem{kamp1960cortical}
A. Kamp, C. W. Sem-Jacobsen, and W. Leeuwen. 
\newblock Cortical responses to modulated light in the human subject.
\newblock \emph{Acta physiologica scandinavica}  48(1), 1-12, 1960.


\bibitem[Kemp et~al.(2002)Kemp, Gray, Eide, Silberstein, and
  Nathan]{kemp2002steady}
A.~Kemp, M.~Gray, P.~Eide, R.~Silberstein, and P.~Nathan.
\newblock Steady-state visually evoked potential topography during processing
  of emotional valence in healthy subjects.
\newblock \emph{NeuroImage}, 17\penalty0 (4):\penalty0 1684--1692, 2002.

\bibitem[Kettenring(1971)]{kettenring1971canonical}
J.~R. Kettenring.
\newblock Canonical analysis of several sets of variables.
\newblock \emph{Biometrika}, 58\penalty0 (3):\penalty0 433--451, 1971.

\bibitem{koskinen2014uncovering}
M.~Koskinen and M~Seppä.
\newblock Uncovering cortical MEG responses to listened audiobook stories.
\newblock \emph{NeuroImage}, 2014.

\bibitem{lin2006frequency}
Z.  Lin, C. Zhang, W. Wu, and X. Gao. (2006). 
\newblock Frequency recognition based on canonical correlation analysis for SSVEP-based BCIs. 
\newblock \emph{IEEE Transactions on Biomedical Engineering}, 53(12), 2610-2614, 2006.

\bibitem[Lankinen et~al.(2014)Lankinen, Saari, Hari, and
  Koskinen]{lankinen2014intersubject}
K.~Lankinen, J.~Saari, R.~Hari, and M.~Koskinen.
\newblock Intersubject consistency of cortical meg signals during movie
  viewing.
\newblock \emph{NeuroImage}, 92:\penalty0 217--224, 2014.

\bibitem[Middendorf et~al.(2000)Middendorf, McMillan, Calhoun, Jones,
  et~al.]{middendorf2000brain}
M.~Middendorf, G.~McMillan, G.~Calhoun, K.~S. Jones, et~al.
\newblock Brain-computer interfaces based on the steady-state visual-evoked
  response.
\newblock \emph{IEEE Transactions on Rehabilitation Engineering}, 8\penalty0
  (2):\penalty0 211--214, 2000.

\bibitem[Ming and Shangkai(1999)]{ming1999eeg}
C.~Ming and G.~Shangkai.
\newblock An eeg-based cursor control system.
\newblock In \emph{[Engineering in Medicine and Biology, 1999. 21st Annual
  Conference and the 1999 Annual Fall Meetring of the Biomedical Engineering
  Society] BMES/EMBS Conference, 1999. Proceedings of the First Joint},
  volume~1, pages 669--vol. IEEE, 1999.

\bibitem[Morgan et~al.(1996)Morgan, Hansen, and Hillyard]{morgan1996selective}
S.~Morgan, J.~Hansen, and S.~Hillyard.
\newblock Selective attention to stimulus location modulates the steady-state
  visual evoked potential.
\newblock \emph{Proceedings of the National Academy of Sciences}, 93\penalty0
  (10):\penalty0 4770--4774, 1996.

\bibitem[M{\"u}ller et~al.(2006)M{\"u}ller, Andersen, Trujillo, Valdes-Sosa,
  Malinowski, and Hillyard]{muller2006feature}
M.~M{\"u}ller, S.~Andersen, N.~Trujillo, P.~Valdes-Sosa, P.~Malinowski, and
  S.~Hillyard.
\newblock Feature-selective attention enhances color signals in early visual
  areas of the human brain.
\newblock \emph{Proceedings of the National Academy of Sciences}, 103\penalty0
  (38):\penalty0 14250--14254, 2006.
  
\bibitem{nam2013common}
Y. Nam, A. Cichocki, and S. Choi. 
\newblock Common spatial patterns for steady-state somatosensory evoked potentials.
\newblock \emph{ Engineering in Medicine and Biology Society (EMBC), 2013 35th Annual International Conference of the IEEE}, 2013.
  
  
\bibitem[Neuenschwander and Flury(1995)]{neuenschwander1995common}
B.~E. Neuenschwander and B.~D. Flury.
\newblock Common canonical variates.
\newblock \emph{Biometrika}, 82\penalty0 (3):\penalty0 553--560, 1995.

\bibitem[Oostenveld and Praamstra(2001)]{oostenveld2001five}
R.~Oostenveld and P.~Praamstra.
\newblock The five percent electrode system for high-resolution eeg and erp
  measurements.
\newblock \emph{Clinical neurophysiology}, 112\penalty0 (4):\penalty0 713--719,
  2001.

\bibitem{pan2011enhancing}
J. Pan, X. Gao, F.  Duan, Z. Yan, and S. Gao.
\newblock Enhancing the classification accuracy of steady-state visual evoked potential-based brain-computer interfaces using phase constrained canonical correlation analysis. 
\newblock \emph{Journal of neural engineering}, 8(3), 036027, 2011.

\bibitem[Parra et~al.(2005)Parra, Spence, Gerson, and Sajda]{parra2005recipes}
L.~C. Parra, C.~D. Spence, A.~D. Gerson, and P.~Sajda.
\newblock Recipes for the linear analysis of eeg.
\newblock \emph{Neuroimage}, 28\penalty0 (2):\penalty0 326--341, 2005.

\bibitem[Pouryazdian and Erfanian(2009)]{pouryazdian2009detection}
S.~Pouryazdian and A.~Erfanian.
\newblock Detection of steady-state visual evoked potentials for brain-computer
  interfaces using pca and high-order statistics.
\newblock In \emph{World Congress on Medical Physics and Biomedical
  Engineering, September 7-12, 2009, Munich, Germany}, pages 480--483.
  Springer, 2009.

\bibitem{regan1966some}
D.~Regan.
\newblock Some characteristics of average steady-state and transient responses evoked by modulated light.
\newblock \emph{ Electroencephalography and clinical neurophysiology} , 20penalty0 (3):\penalty0 238--248 1966

\bibitem[Regan(1989)]{regan1989human}
D.~Regan.
\newblock Human brain electrophysiology: evoked potentials and evoked magnetic
  fields in science and medicine.
\newblock 1989.

\bibitem[Riera et~al.(2012)Riera, Ogawa, Goto, Sumiyoshi, Nonaka, Evans,
  Miyakawa, and Kawashima]{riera2012pitfalls}
J.~J. Riera, T.~Ogawa, T.~Goto, A.~Sumiyoshi, H.~Nonaka, A.~Evans, H.~Miyakawa,
  and R.~Kawashima.
\newblock Pitfalls in the dipolar model for the neocortical eeg sources.
\newblock \emph{Journal of neurophysiology}, 108\penalty0 (4):\penalty0
  956--975, 2012.

\bibitem[Schurger et~al.(2010)Schurger, Pereira, Treisman, and
  Cohen]{schurger2010reproducibility}
A.~Schurger, F.~Pereira, A.~Treisman, and J.~D. Cohen.
\newblock Reproducibility distinguishes conscious from nonconscious neural
  representations.
\newblock \emph{Science}, 327\penalty0 (5961):\penalty0 97--99, 2010.

\bibitem[Silberstein(1995)]{Silberstein1995steady}
R.~B. Silberstein.
\newblock Steady-state visually evoked potentials, brain resonances, and
  cognitive processes.
\newblock In P.~L. Nunez, editor, \emph{Neocortical dynamics and human brain
  rhythms}. Oxford University Press, Oxford, 1995.

\bibitem[Silberstein et~al.(1995)Silberstein, Pipingas,
  et~al.]{silberstein1995steadyb}
R.~B. Silberstein, A.~Pipingas, et~al.
\newblock Steady-state visually evoked potential topography during the
  wisconsin card sorting test.
\newblock \emph{Electroencephalography and Clinical Neurophysiology/Evoked
  Potentials Section}, 96\penalty0 (1):\penalty0 24--35, 1995.

\bibitem[Srinivasan and Petrovic(2006)]{srinivasan2006meg}
R.~Srinivasan and S.~Petrovic.
\newblock Meg phase follows conscious perception during binocular rivalry
  induced by visual stream segregation.
\newblock \emph{Cerebral Cortex}, 16\penalty0 (5):\penalty0 597--608, 2006.

\bibitem{tweel1965human}
L. H. Van der Tweel, and H. F. E. Verduyn Lunel. 
\newblock Human visual responses to sinusoidally modulated light.
\newblock \emph{Electroencephalography and Clinical Neurophysiology} 18.6: 587-598, 1965.

\bibitem{tweel1969signal}
L. H. van der Tweel LH and H Spekreijse.
\newblock  Signal transport and rectification in the human evoked-response system.
\newblock \emph{Ann N Y Acad Sci.} 156(2):678-95. 1969.

\bibitem[Vialatte et~al.(2010)Vialatte, Maurice, Dauwels, and
  Cichocki]{vialatte2010steady}
F.-B. Vialatte, M.~Maurice, J.~Dauwels, and A.~Cichocki.
\newblock Steady-state visually evoked potentials: focus on essential paradigms
  and future perspectives.
\newblock \emph{Progress in neurobiology}, 90\penalty0 (4):\penalty0 418--438,
  2010.

\bibitem[Wang et~al.(2006)Wang, Wang, Gao, Hong, and Gao]{wang2006practical}
Y.~Wang, R.~Wang, X.~Gao, B.~Hong, and S.~Gao.
\newblock A practical vep-based brain-computer interface.
\newblock \emph{IEEE Transactions on Neural Systems and Rehabilitation Engineering}, 14\penalty0 (2):\penalty0 234--240, 2006.

\bibitem{wang2008brain}
Y. Wang, X., Gao, B. Hong, C. Jia, and S. Gao, S. (2008). 
\newblock Brain-computer interfaces based on visual evoked potentials. 
\newblock \emph{IEEE Engineering in Medicine and Biology Magazine}, 27(5), 64-71, 2008.

\bibitem[Wolters et~al.(2006)Wolters, Anwander, Tricoche, Weinstein, Koch, and
  MacLeod]{wolters2006influence}
C.~Wolters, A.~Anwander, X.~Tricoche, D.~Weinstein, M.~Koch, and R.~MacLeod.
\newblock Influence of tissue conductivity anisotropy on eeg/meg field and
  return current computation in a realistic head model: a simulation and
  visualization study using high-resolution finite element modeling.
\newblock \emph{NeuroImage}, 30\penalty0 (3):\penalty0 813--826, 2006.

\bibitem{zhang2011multiway}
Y. Zhang, G. Zhou, Q. Zhao, A. Onishi, J. Jin, X. Wang, and A. Cichocki, A. (2011, January). 
\newblock Multiway canonical correlation analysis for frequency components recognition in SSVEP-based BCIs. 
\newblock In \emph{Proceedings Neural Information Processing}, 287-295, 2011.

\bibitem{zhang2013l1}
Y. Zhang, G. Zhou, J. Jin, M. Wang, X. Wang, and A. Cichocki.
\newblock L1-regularized multiway canonical correlation analysis for SSVEP-based BCI. 
\newblock \emph{IEEE Transactions on Neural Systems and Rehabilitation Engineering}, 21(6):887-897, 2013. 

\bibitem{zhu2010survey}
D. Zhu, J. Bieger, G. G.  Molina, and R. M. Aarts 
\newblock A survey of stimulation methods used in SSVEP-based BCIs. 
\newblock \emph{Computational intelligence and neuroscience}, 2010:1, 2010.

\end{thebibliography}


\end{document}